\documentclass[aps,prd,twocolumn,superscriptaddress,floatfix]{revtex4-2}
\usepackage{amsmath,amssymb}
\usepackage{xcolor}

\usepackage[colorlinks=true
  ,urlcolor=blue
  ,anchorcolor=blue
  ,citecolor=blue
  ,filecolor=blue
  ,linkcolor=blue
  ,menucolor=blue
  ,linktocpage=true
  ,pdfproducer=medialab
  ,pdfa=true
]{hyperref}
\usepackage{orcidlink}
\usepackage{multirow} 
\begin{document}
\raggedbottom
% ---------------------------------------
%\title{From Proto- to Cold Strange Quark Stars Admixed with Fermionic Dark Matter}
\title{Early Evolution of Strange Quark Stars Admixed with Fermionic Dark Matter}
%\title{Dark Matter in Strange Quark Stars From Birth to the Cold Stage}
\author{Rajasmita Sahoo\,\orcidlink{0000-0002-9265-5025}}
\email{rsphysics58@gmail.com}
\affiliation{National Institute of Technology, Tiruchirappalli, Tamil Nadu-620015, India}
\affiliation{Institute of Physics, Sachivalaya Marg, Bhubaneswar, Odisha-751005, India}
\author{Mrutunjaya Bhuyan\,\orcidlink{0000-0002-8677-4220}}
\email{mrutunjaya.b@iopb.res.in}
\affiliation{Institute of Physics, Sachivalaya Marg, Bhubaneswar, Odisha-751005, India}
\affiliation{Homi Bhabha National Institute, Training School Complex, Anushakti Nagar, Mumbai 400094, India}
\date{\today}
% ------------------------------------
\begin{abstract}
We study strange quark stars admixed with fermionic dark matter (DM) in three quasistatic stages of their early evolution: the neutrino-trapped stage, the heated stage after deleptonization, and the cold stage. Quark matter is described by the MIT bag model at finite temperature with thermal gluons, a repulsive vector interaction, and, in the first stage, trapped neutrinos. The DM is modeled as a cold, asymmetric gas of fermions with masses $m_\chi = 0.3$, $1$, and $3$~GeV, with and without a repulsive self-interaction, and interacts with the quark matter only through gravity. Using the two-fluid Tolman--Oppenheimer--Volkoff equations, we compute the structure, the maximum masses, and the tidal deformability of the stars in each stage for DM mass fractions $F_D$ up to $5\%$, and we follow individual stars at fixed numbers of baryons and DM particles. The effect of DM is nearly the same in the three stages: for $m_\chi = 1$~GeV and $F_D = 5\%$, DM lowers the maximum mass by about $8\%$ and the tidal deformability of a $1.4\,M_\odot$ star by about $26\%$ in each stage, whereas the stage itself changes these quantities by only a few percent. Heavy DM forms a compact core; for $m_\chi = 3$~GeV without self-interaction, the mass limit of this core can set the maximum mass of the star. Light DM can form an extended halo, which does not increase the maximum mass for the fractions considered. For $m_\chi = 1$~GeV and $F_D = 5\%$, DM lowers the largest baryon mass of a newborn star by about $0.23\,M_\odot$, but this limit increases from the neutrino-trapped to the cold stage in all cases considered, so that DM does not cause a delayed collapse to a black hole. In our model, pulsars of about $2\,M_\odot$ require a stiff quark equation of state and restrict the DM fraction to a few percent unless the DM is light.
\end{abstract}
% -----------------------------------
\maketitle
% ------------------------------
\section{Introduction}
\label{sec:intro}
The existence of dark matter (DM) is inferred from its gravitational effects on galactic rotation curves, the dynamics of galaxy clusters, gravitational lensing, and anisotropies in the cosmic microwave background~\cite{1970ApJ...159..379R,1980ApJ...238..471R,Zwicky:1933gu,Clowe:2006eq,Bertone:2004pz,Planck:2018vyg}. Together, these observations indicate that most of the matter in the Universe is nonbaryonic. Within the standard $\Lambda$CDM model, Planck measurements imply that DM accounts for approximately $84\%$ of the matter content~\cite{Planck:2018vyg}. The particle nature of DM, however, remains unknown. Direct-detection experiments~\cite{LZ:2022lsv,LZ:2024zvo,XENON:2023cxc,PandaX-4T:2021bab}, indirect searches~\cite{Fermi-LAT:2015att}, and collider experiments~\cite{Boveia:2018yeb} have not yet yielded a conclusive DM signal, although they have placed increasingly stringent constraints on several classes of DM models~\cite{PhysRevD.110.030001}. Conventional searches based on nuclear recoils become less sensitive for DM masses below a few GeV because the energy transferred to the target nucleus may fall below the detector threshold~\cite{Billard:2021uyg}. This loss of sensitivity motivates complementary probes of DM based on its gravitational effects in dense astrophysical systems.
% --------------------------

Compact stars are especially useful for this purpose. Their central densities can reach several times the nuclear saturation density, and their gravitational fields are strong, so that even a small additional component contributes to the total stress-energy and alters the hydrostatic balance appreciably. If DM scatters from stellar matter and loses enough kinetic energy to become gravitationally bound, compact stars can capture it from the Galactic halo over their lifetimes~\cite{Goldman:1989nd,Kouvaris:2007ay,Bertone:2007ae}. Under conditions typical of the solar neighborhood, capture estimates for compact stars generally yield a very small accumulated mass fraction~\cite{DelPopolo:2019nng,DelPopolo:2020hel,Ivanytskyi:2019wxd}. An appreciable DM component at birth would require either prior accumulation by the progenitor or the production or capture of DM during stellar collapse~\cite{Ivanytskyi:2019wxd,Nelson:2018xtr}. If the nongravitational interaction between DM and ordinary matter is too weak to affect the stellar structure, the star can be described by two fluids coupled only through the common spacetime geometry, and any DM self-interaction acts only within the dark sector~\cite{Sandin:2008db,Leung:2011zz}. Even a DM fraction of a few percent can appreciably modify the mass--radius relation and the tidal deformability of a compact star~\cite{Sandin:2008db,Ellis:2018bkr}. Because the amount of DM in a compact star is not known, we treat the global DM mass fraction $F_D$ as a free parameter, as is common in two-fluid studies~\cite{Grippa:2024ach}. Relevant observational benchmarks include the masses of the most massive pulsars, measured through radio timing~\cite{Demorest:2010bx,Antoniadis:2013pzd,NANOGrav:2019jur,Fonseca:2021wxt}, the simultaneous mass and radius constraints from NICER~\cite{Riley:2019yda,Miller:2019cac,Riley:2021pdl,Miller:2021qha}, and the tidal deformability inferred from GW170817~\cite{LIGOScientific:2017vwq,LIGOScientific:2018cki}. The impact of DM depends not only on its abundance and microphysics but also on the equation of state (EOS) and thermodynamic state of the ordinary matter.
% --------------------------
 
The EOS of the ordinary matter, in turn, depends on its composition at densities above nuclear saturation. In addition to neutron stars, whose interiors are expected to be predominantly hadronic, there may exist compact stars that consist of deconfined quark matter throughout. According to the Bodmer--Witten hypothesis, matter composed of up, down, and strange quarks, known as strange quark matter, could have a lower energy per baryon than the most stable atomic nuclei, such as $^{56}$Fe, at zero pressure and might thus be the true ground state of strongly interacting matter~\cite{1971PhRvD...4.1601B,1984PhRvD..30..272W}. Such matter would be self-bound, and stars composed of it are known as strange quark stars, or simply strange stars~\cite{Farhi:1984qu,1986ApJ...310..261A,1986A&A...160..121H}. The matter is bound by the strong interaction rather than by gravity, and in the idealized bare-star picture the star has a sharp surface at nonzero density. At sufficiently small masses, the density is nearly uniform and $M\propto R^3$, so that light strange stars are small. Neutron stars, in contrast, are bound by gravity, and at low masses their radius increases as the mass decreases. Strange quark stars have been proposed as one possible interpretation of unusually compact objects, including the central compact object in the supernova remnant HESS~J1731$-$347, with $M = 0.77^{+0.20}_{-0.17}\,M_\odot$ and $R = 10.4^{+0.86}_{-0.78}$~km~\cite{2022NatAs...6.1444D,DiClemente:2022wqp,Horvath:2023uwl,Oikonomou:2023otn}, although current observations do not determine its composition. Any viable model of strange quark stars must also support the most massive observed pulsars, with masses close to $2\,M_\odot$~\cite{Demorest:2010bx,Antoniadis:2013pzd,NANOGrav:2019jur,Fonseca:2021wxt}. In the conventional MIT bag model, parameter sets consistent with the stability of strange quark matter can yield maximum masses below this value. This motivates extensions with a repulsive vector interaction among quarks, which stiffens the EOS and increases the maximum mass, while it also raises the energy per baryon and thus narrows the stability window~\cite{Lopes:2020btp,Lopes:2020dvs}.
% -----------------------

Core-collapse supernovae leave behind compact stars that are initially hot and rich in leptons~\cite{Burrows:1986me,Prakash:1996xs,Pons:1998mm}. During the first few seconds after formation, the interior temperature can reach several tens of MeV, and neutrinos remain trapped because their mean free paths are much shorter than the stellar radius. Over the following tens of seconds, neutrino diffusion carries away lepton number and energy, a process known as deleptonization. The temperature does not decrease monotonically: during deleptonization the interior heats up, and the entropy per baryon increases from about 1 to 2~\cite{Prakash:1996xs,Pons:1998mm}. The star subsequently cools through neutrino emission and approaches a cold state transparent to neutrinos. The neutrino burst from SN~1987A, observed over roughly ten seconds, provided direct evidence that newly formed compact stars lose most of their binding energy through neutrinos~\cite{Kamiokande-II:1987idp,PhysRevLett.58.1494}. Because the hydrodynamic timescale, $\sim 10^{-4}$~s, is much shorter than the neutrino diffusion and cooling times, the evolution can be approximated as a sequence of quasistatic configurations in hydrostatic equilibrium~\cite{Prakash:1996xs,Pons:1998mm}. Comparing these configurations at fixed total baryon number follows the changes in the mass and radius of the same star and shows whether a star stable at birth can later become unstable and undergo a delayed collapse to a black hole~\cite{Prakash:1996xs}. For a star containing DM, the number of DM particles must also be held fixed in such a comparison, rather than the DM mass fraction. Studies of DM-admixed proto-neutron stars have treated the DM either as part of a single fluid in thermal equilibrium with the nucleons through a Higgs-portal coupling~\cite{Issifu:2025qqw} or as a separate fluid whose mass fraction is kept fixed during the evolution~\cite{Issifu:2025jac,Issifu:2025gsq}, and a treatment that conserves the number of DM particles separately has been left for future work~\cite{Issifu:2025jac}. Calculations for newborn strange quark stars, on the other hand, have been performed without a DM component~\cite{Dexheimer:2013eua,Lopes:2020dvs,Issifu:2023qoo,Chen:2026nxv}. 
% ----------------------------------------

In the two-fluid model, each fluid has its own EOS and conservation law, so that the numbers of baryons and DM particles are conserved separately~\cite{Sandin:2008db,Ciarcelluti:2010ji,Leung:2011zz}. At fixed global DM fraction $F_D$, increasing the DM particle mass generally makes the DM distribution more compact, while a repulsive self-interaction makes it more extended: sufficiently heavy DM forms a central core, whereas sufficiently light DM can form an extended halo around the quark component~\cite{Nelson:2018xtr,Ivanytskyi:2019wxd}. Existing studies of DM-admixed strange stars have considered only cold configurations~\cite{Mukhopadhyay:2015xhs,Panotopoulos:2017pgv,PhysRevD.98.083001,PhysRevD.104.083016,Jimenez:2021nmr,Ferreira:2022fjo,Lopes:2023uxi,Marzola:2024ame,Yang:2024vjx}. To the best of our knowledge, the early evolution of DM-admixed strange quark stars has not yet been studied. In particular, no study has followed such a star at fixed numbers of baryons and DM particles from the neutrino-trapped stage through the heated stage to the cold stage, using a finite-temperature quark EOS and a cold fermionic DM fluid coupled to the quark matter only through gravity.
% -----------------------------------------

The central question addressed in this work is whether a DM component changes the stability of a newly formed strange quark star as it evolves from the hot, neutrino-trapped stage toward the cold stage. In proto-neutron stars that contain hyperons, a kaon condensate, or quark matter, the loss of the trapped neutrinos lowers the threshold densities of these components and softens the EOS, so that the maximum mass decreases during deleptonization and a delayed collapse to a black hole becomes possible~\cite{Prakash:1996xs}. Strange quark matter, in contrast, already contains all three light flavors, is self-bound, and remains strongly degenerate in the hot stages, so that no new components appear when the neutrinos escape. A cold DM component that couples to the quark matter only through gravity lowers the stability limit of the star, but it can affect how this limit evolves only through the stage dependence of the quark EOS or, for a DM core close to its own mass limit, through the gravitational field of the quark matter in which the core is embedded. Following individual stars also shows how the DM distribution responds to the evolution of the quark matter, which differs for a compact DM core and an extended DM halo. Our principal objective is therefore to determine how the DM fraction, the DM particle mass, its self-interaction, and the stiffness of the quark matter affect the largest baryon mass, and whether the evolution permits a delayed collapse within the adopted model.
% -----------------------------

In this work, we study strange quark stars containing fermionic DM in three quasistatic stages: (i) a neutrino-trapped stage with entropy per baryon $s_Q/n_B=1$ and lepton fraction $Y_l=0.4$; (ii) a heated, deleptonized stage with $s_Q/n_B=2$ and $\mu_\nu=0$; and (iii) a cold stage with $T=0$. As in earlier studies of proto-strange stars~\cite{Dexheimer:2013eua,Lopes:2020dvs,Issifu:2023qoo}, we assume that the star consists of strange quark matter from the neutrino-trapped stage on; a possible formation scenario is discussed in Ref.~\cite{DiClemente:2022wqp}. Quark matter is described by the MIT bag model at finite temperature with thermal gluons, a repulsive vector interaction, and, in the first stage, trapped neutrinos. We take $B^{1/4}=150$~MeV and $0\leq G_V\leq0.15~\mathrm{fm}^2$, for which strange quark matter is absolutely stable (Sec.~\ref{subsec:parameters}). The DM is treated as a cold, asymmetric gas of spin-$1/2$ fermions with masses $m_\chi=0.3$, $1$, and $3$~GeV and repulsive self-interaction strengths $G_\chi=0$, $0.05$, and $0.20~\mathrm{fm}^2$. It is nonannihilating, does not couple directly to neutrinos, and does not exchange heat with the quark matter; these are assumptions of the model rather than general properties of DM. We consider $F_D \leq 5\%$ and solve the two-fluid Tolman--Oppenheimer--Volkoff equations. For each stage, we calculate the mass--radius relation, the maximum gravitational mass, the tidal deformability, and the DM distribution, in particular whether the DM forms a core or an extended halo. We also follow individual stars through the three stages at fixed numbers of baryons and DM particles. The DM lowers the maximum mass and the tidal deformability by nearly the same fraction in the three stages, whereas the stage itself changes these quantities by only a few percent. DM also lowers the largest baryon mass of a newborn star, but in all cases considered this limit increases from the neutrino-trapped to the cold stage. Because the DM is cold and its EOS is the same in all stages, it lowers the stability limits of the three stages by similar amounts, so that their ordering is set mainly by the quark matter. Consequently, a DM-admixed strange star that is stable at birth remains stable during its early evolution, and DM does not cause a delayed collapse to a black hole.
% --------------------------------------

The paper is organized as follows. In Sec.~\ref{sec:eos}, we present the EOSs of strange quark matter at finite temperature and of fermionic DM, together with the conditions that define the three evolutionary stages. In Sec.~\ref{sec:structure}, we introduce the two-fluid structure equations, the fixed particle numbers used to follow an individual star, and the tidal deformability. Our results are presented and discussed in Sec.~\ref{sec:results}, and our conclusions are given in Sec.~\ref{sec:conclusions}. 
%======================================================================
\section{Equations of state}
\label{sec:eos}
\subsection{Strange quark matter at finite temperature}
\label{subsec:sqm}
% --------------------------------------------------- 

We describe strange quark matter (SQM) by the MIT bag model~\cite{PhysRevD.9.3471,Farhi:1984qu} at finite temperature, following earlier studies of proto-quark stars~\cite{Dexheimer:2013eua,Lopes:2020dvs}. Quark matter is treated as a relativistic Fermi gas of quarks and, at $T>0$, antiquarks, confined by a constant bag pressure $B$; perturbative QCD corrections are neglected ($\alpha_s = 0$). The quarks coexist with electrons and, at $T>0$, positrons, and the total mixture is required to be electrically neutral; muons are neglected. In the neutrino-trapped stage, the mixture also contains electron neutrinos and antineutrinos, and at nonzero temperature thermal gluons contribute to the pressure, energy density, and entropy. A repulsive vector interaction among the quarks, introduced in Sec.~\ref{subsec:vector}, stiffens the EOS. We take $m_u = 2.2$~MeV, $m_d = 4.8$~MeV, and $m_s = 95$~MeV, close to the values of the Particle Data Group~\cite{PhysRevD.110.030001}, and $m_e = 0.511$~MeV. In the description of the matter, we use natural units, $\hbar = c = k_B = 1$.
% ------------------------------ 

For $T > 0$, in the grand canonical ensemble, each fermion species $i$ ($i = u, d, s, e$) with degeneracy $g_i$, mass $m_i$, and effective chemical potential $\mu_i^\ast$ contributes a net number density, energy density, pressure, and entropy density~\cite{Kapusta:2007xjq,Lopes:2020dvs}
\begin{align}
n_i &= \frac{g_i}{2\pi^2}\int_0^\infty p^2
       \left[ f_i(p) - \bar{f}_i(p) \right] dp ,
       \label{eq:n_i}\\
\varepsilon_i &= \frac{g_i}{2\pi^2}\int_0^\infty p^2 E_i
       \left[ f_i(p) + \bar{f}_i(p) \right] dp ,
       \label{eq:e_i}\\
P_i &= \frac{g_i T}{2\pi^2}\int_0^\infty p^2
       \Big\{ \ln\!\left[1 + e^{-(E_i-\mu_i^\ast)/T}\right] \nonumber\\
      &\qquad\qquad\qquad\quad + \ln\!\left[1 + e^{-(E_i+\mu_i^\ast)/T}\right] \Big\}\, dp ,
       \label{eq:P_i}\\
s_i &= \frac{\varepsilon_i + P_i - \mu_i^\ast n_i}{T} ,
       \label{eq:s_i}
\end{align}
where $E_i = (p^2 + m_i^2)^{1/2}$ and
\begin{equation}
f_i(p) = \frac{1}{e^{(E_i-\mu_i^\ast)/T} + 1} , \quad
\bar{f}_i(p) = \frac{1}{e^{(E_i+\mu_i^\ast)/T} + 1}
\label{eq:FD}
\end{equation}
are the Fermi--Dirac distributions of particles and antiparticles. For the quarks, $\mu_i^\ast$ differs from the thermodynamic chemical potential $\mu_i$ by the vector shift of Eq.~(\ref{eq:mu_shift}); for electrons, $\mu_e^\ast = \mu_e$. Equations~(\ref{eq:n_i})--(\ref{eq:s_i}) give the kinetic contributions; the vector-interaction terms are added in Sec.~\ref{subsec:vector}. The degeneracy factors are $g_q = 6$ for each quark flavor (three colors and two spin states) and $g_e = 2$ for electrons (two spin states). For the electrons, Eq.~(\ref{eq:n_i}) gives the net density $n_e = n_{e^-} - n_{e^+}$. At $T = 0$, for $\mu_i^\ast > m_i$, the antiparticle contributions vanish and the integrals are cut off at the Fermi momentum $k_{F,i} = [(\mu_i^\ast)^2 - m_i^2]^{1/2}$; a species is absent when $\mu_i^\ast \le m_i$.
% ------------------------------------

In the neutrino-trapped stage, electron neutrinos and antineutrinos are treated as massless, with one helicity state for each particle and antiparticle ($g_\nu = 1$); thermal muon and tau neutrinos are neglected because their contribution is small. Their net number density, pressure, and energy density are~\cite{Kapusta:2007xjq}
\begin{gather}
n_\nu = \frac{\mu_\nu^3}{6\pi^2} + \frac{\mu_\nu T^2}{6} ,
\nonumber\\
P_\nu = \frac{7\pi^2 T^4}{360} + \frac{\mu_\nu^2 T^2}{12} + \frac{\mu_\nu^4}{24\pi^2} ,
\qquad
\varepsilon_\nu = 3 P_\nu ,
\label{eq:nu}
\end{gather}
and the corresponding entropy density is $s_\nu = (\varepsilon_\nu + P_\nu - \mu_\nu n_\nu)/T$. In the heated and cold stages, the neutrinos are assumed to have escaped: we set $\mu_\nu = 0$ in the equilibrium conditions and omit the neutrino contribution entirely, including thermal neutrino--antineutrino pairs, so that $n_\nu = P_\nu = \varepsilon_\nu = s_\nu = 0$~\cite{Prakash:1996xs,Dexheimer:2013eua}.
% ------------------------------------

For $T > 0$, the thermal gluons form a free, massless Bose gas with $g_g = 16$, corresponding to eight color degrees of freedom and two physical polarization states, and with vanishing chemical potential~\cite{Kapusta:2007xjq}:
\begin{gather}
P_g = \frac{g_g \pi^2}{90}\,T^4 = \frac{8\pi^2}{45}\,T^4 ,
\nonumber\\
\varepsilon_g = 3 P_g , \qquad
s_g = \frac{32\pi^2}{45}\,T^3 .
\label{eq:gluon}
\end{gather}
% ---------------------------------

\subsection{Vector interaction and total quark-matter EOS}
\label{subsec:vector}

The repulsive interaction among the quarks is mediated by a massive vector field $V_\mu$~\cite{Lopes:2020btp,Lopes:2020dvs}, which we take to couple with the same (universal) strength $g_v$ to all three flavors. In the mean-field approximation for static, uniform matter, only the time component of the field survives, $\langle V_0 \rangle = (g_v/m_v^2)\, n_q$, where $m_v$ is the mass of the vector field and $n_q = n_u + n_d + n_s = 3 n_B$ is the total quark number density, with $n_B$ the baryon number density. The interaction contributes equally to the pressure and to the energy density,
\begin{equation}
P_V = \varepsilon_V = \frac{1}{2}\, G_V\, n_q^2 , \qquad
G_V = \left(\frac{g_v}{m_v}\right)^2 ,
\label{eq:vector}
\end{equation}
and it shifts the chemical potential of every quark flavor by the same amount,
\begin{equation}
\mu_i = \mu_i^\ast + G_V\, n_q , \qquad i = u, d, s ,
\label{eq:mu_shift}
\end{equation}
where $\mu_i$ is the thermodynamic chemical potential and $\mu_i^\ast$ the effective chemical potential that enters the Fermi--Dirac distributions in Eqs.~(\ref{eq:n_i})--(\ref{eq:FD}). In this approximation the vector term does not depend on temperature and does not contribute to the entropy. Only the combination $G_V$ enters the EOS, so that $g_v$ and $m_v$ need not be specified separately, and we treat $G_V$ as a parameter that controls the stiffness of the EOS. The total pressure, energy density, and entropy density of the quark-matter fluid, which includes the leptons and gluons, are
\begin{align}
P_Q &= \sum_{i} P_i + P_\nu + P_g + P_V - B ,
\label{eq:PQ}\\
\varepsilon_Q &= \sum_{i} \varepsilon_i + \varepsilon_\nu + \varepsilon_g + \varepsilon_V + B ,
\label{eq:eQ}\\
s_Q &= \sum_{i} s_i + s_\nu + s_g ,
\label{eq:sQ}
\end{align}
where the sums run over $i = u, d, s, e$, the neutrino terms are present only in the neutrino-trapped stage, and the gluon terms only for $T > 0$. With the full chemical potentials of Eq.~(\ref{eq:mu_shift}), these expressions satisfy the thermodynamic identity $\varepsilon_Q + P_Q = T s_Q + \sum_j \mu_j n_j$, where $j$ runs over all quarks and leptons. The bag constant lowers the pressure and raises the energy density, so that $P_Q$ vanishes at a finite baryon density $n_{B,0}$, which defines the surface of the self-bound star in each stage.
% ------------------------
% FIGURE -01
\begin{figure}[tbp]
\centering
\includegraphics[width=\columnwidth]{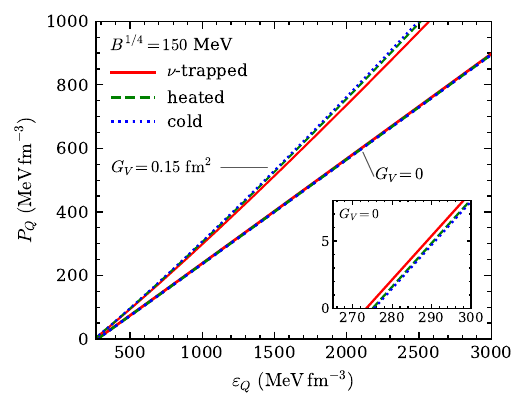}
\caption{Pressure $P_Q$ versus energy density $\varepsilon_Q$ of SQM for $B^{1/4} = 150$~MeV in the neutrino-trapped,
heated, and cold stages (Table~\ref{tab:stages}). The lower group of curves is for $G_V = 0$ and
the upper group for $G_V = 0.15$~fm$^2$. The inset magnifies the region near the stellar surface
($P_Q = 0$) for $G_V = 0$.}
\label{fig:P_eps}
\end{figure}
% ---------------------------------------
\subsection{Equilibrium conditions}
\label{subsec:equilibrium}
Quark matter is in $\beta$ equilibrium under the weak processes $d \to u + e^- + \bar{\nu}_e$, $s \to u + e^- + \bar{\nu}_e$, $u + e^- \to d + \nu_e$, $u + e^- \to s + \nu_e$, and $u + d \leftrightarrow u + s$, which imply
\begin{equation}
\mu_d = \mu_s = \mu_u + \mu_e - \mu_\nu ,
\label{eq:beta}
\end{equation}
with $\mu_\nu = 0$ in the heated and cold stages. Since the vector shift of Eq.~(\ref{eq:mu_shift}) is the same for all flavors, it cancels in Eq.~(\ref{eq:beta}), which then holds equally for the effective chemical potentials $\mu_i^\ast$. As a consequence, at a given baryon density the composition and the temperature of the matter do not depend on $G_V$. The matter is electrically neutral, and its baryon density is given by the quark densities,
\begin{equation}
\frac{2}{3}\, n_u - \frac{1}{3}\, n_d - \frac{1}{3}\, n_s - n_e = 0 ,
\quad
n_B = \frac{1}{3}\left( n_u + n_d + n_s \right) .
\label{eq:neutrality}
\end{equation}
In each hot stage, the entropy per baryon and, while neutrinos are trapped, the lepton fraction are uniform throughout the star,
\begin{equation}
\frac{s_Q}{n_B} = \mathrm{const} , \qquad
Y_l = \frac{n_e + n_\nu}{n_B} = \mathrm{const} ,
\label{eq:SYl}
\end{equation}
with the values given in Sec.~\ref{subsec:stages}. At each baryon density we solve Eqs.~(\ref{eq:beta})--(\ref{eq:SYl}) numerically for $(\mu_d^\ast, \mu_e, \mu_\nu, T)$ in the neutrino-trapped stage, $(\mu_d^\ast, \mu_e, T)$ in the heated stage, and $(\mu_d^\ast, \mu_e)$ in the cold stage. Muons are neglected because their chemical potential, $\mu_\mu = \mu_e - \mu_\nu$, satisfies $|\mu_\mu| < 70$~MeV in the stars, below the muon mass: we estimate that muons and antimuons, including thermal pairs, amount to less than $10^{-3}$ of the baryon density, so that their contribution to the EOS is negligible.
% --------------------------------
\subsection{Evolutionary stages}
\label{subsec:stages}
As discussed in Sec.~\ref{sec:intro}, we describe the early evolution of the star as a sequence of quasistatic configurations in hydrostatic equilibrium~\cite{Prakash:1996xs,Dexheimer:2013eua}. Following Ref.~\cite{Prakash:1996xs}, we consider three stages, summarized in Table~\ref{tab:stages}: (i)~the neutrino-trapped stage of the newborn star, with $s_Q/n_B = 1$ and $Y_l = 0.4$; (ii)~the heated stage, with $s_Q/n_B = 2$, reached after deleptonization has heated the matter and the neutrinos have escaped (Sec.~\ref{subsec:sqm}); and (iii)~the cold, catalyzed stage with $T = 0$, reached after the star has cooled by neutrino emission. Because the entropy per baryon rather than the temperature is fixed, the temperature is determined at each density by Eq.~(\ref{eq:SYl}); it increases with density and hence from the surface toward the center of the star. The DM is cold in all three stages (Sec.~\ref{subsec:dm}). 
% ---------------------------------
% FIGURE - 02
\begin{figure}[tbp]
\centering
\includegraphics[width=\columnwidth]{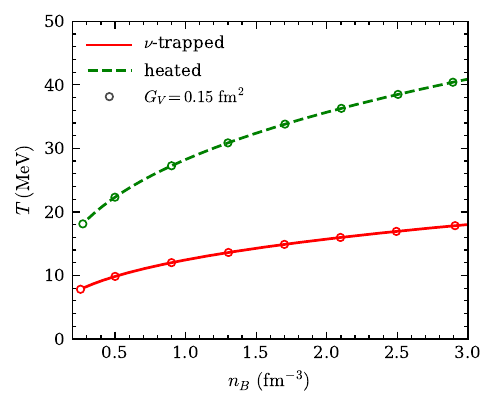}
\caption{Temperature of SQM along the isentropes of the neutrino-trapped ($s_Q/n_B = 1$, $Y_l = 0.4$) and heated ($s_Q/n_B = 2$) stages as a function of the baryon density. Lines are for $G_V = 0$ and open circles for $G_V = 0.15$~fm$^2$; the first circle of each set marks the surface density for $G_V = 0.15$~fm$^2$.}
\label{fig:T_nB}
\end{figure}
% ------------------------
% TABLE - 01
\begin{table}[t]
\caption{\label{tab:stages}Thermodynamic conditions of quark matter in the three evolutionary
stages, and the unknowns solved for at each baryon density. Ellipses indicate quantities that are not fixed but follow from the equilibrium conditions.}
\setlength{\tabcolsep}{4pt}
\begin{ruledtabular}
\begin{tabular}{llll}
 & $\nu$-trapped & Heated & Cold \\
\hline
$s_Q/n_B$ & $1$ & $2$ & $0$ \\
$Y_l$ & $0.4$ & $\cdots$ & $\cdots$ \\
$T$ & from $s_Q/n_B$ & from $s_Q/n_B$ & $0$ \\
$\mu_\nu$ & from $Y_l$ & $0$ & $0$ \\
Trapped $\nu_e$, $\bar\nu_e$ & yes & no & no \\
Antiparticles & yes & yes & no \\
Thermal gluons & yes & yes & no \\
Unknowns & $\mu_d^\ast, \mu_e, \mu_\nu, T$ & $\mu_d^\ast, \mu_e, T$ & $\mu_d^\ast, \mu_e$ \\
\end{tabular}
\end{ruledtabular}
\end{table} 
% --------------------------
\subsection{Parameters and stability window}
\label{subsec:parameters}
We take $B^{1/4} = 150$~MeV and four values of the vector coupling, $G_V = 0$, $0.05$, $0.10$, and $0.15$~fm$^2$. These parameters are constrained by the requirement that SQM be absolutely stable~\cite{Farhi:1984qu}: the energy per baryon of cold three-flavor matter at zero pressure must lie below that of the most stable nuclei, $E/A(uds) < 930$~MeV, while two-flavor ($ud$) matter must be unstable, $E/A(ud) > 930$~MeV, so that ordinary nuclei do not decay into deconfined $u$--$d$ matter. If surface effects in nuclei are taken into account, the latter bound becomes $E/A(ud) > 934$~MeV~\cite{Farhi:1984qu}. The values for our parameter sets, listed in Table~\ref{tab:stability}, satisfy both bounds. The vector interaction raises $E/A$ for both compositions and thus narrows the stability window; for $G_V = 0.15$~fm$^2$ it is $140.0 < B^{1/4} < 152.0$~MeV (with the bound $E/A(ud) > 930$~MeV), so that $B^{1/4} = 150$~MeV lies inside the window for all values of $G_V$ considered. Since these conditions refer to cold matter, they are not affected by the thermal gluon contribution. The EOS is causal in all stages, with $c_s^2 = dP/d\varepsilon \le 0.55$ at the densities reached in the stars.
% -----------------------------
% TABLE - 02
\begin{table}[t]
\caption{Baryon density $n_{B,0}$ of cold three-flavor ($uds$) quark matter at zero pressure, and the energy per baryon at zero pressure for cold three-flavor and two-flavor ($ud$) quark matter, for $B^{1/4} = 150$~MeV and the four values of $G_V$. Absolute stability of SQM requires $E/A(uds) < 930$~MeV~$< E/A(ud)$, which is satisfied in all cases, also with the stricter bound $E/A(ud) > 934$~MeV.}
\label{tab:stability}
\begin{ruledtabular}
\begin{tabular}{cccc}
$G_V$ & $n_{B,0}$ & $E/A(uds)$ & $E/A(ud)$ \\
(fm$^2$) & (fm$^{-3}$) & (MeV) & (MeV) \\
\hline
0.00 & 0.314 & 878.1 & 966.2 \\
0.05 & 0.299 & 891.7 & 978.1 \\
0.10 & 0.287 & 904.7 & 989.6 \\
0.15 & 0.277 & 917.2 & 1000.8 \\
\end{tabular}
\end{ruledtabular}
\end{table} 
% -------------------------------------- 
\subsection{Fermionic dark matter}
\label{subsec:dm}
The second component of the star is DM, which we model as a single species of spin-$1/2$ fermions $\chi$ of mass $m_\chi$. We take the DM to be asymmetric, i.e., composed of particles only, so that it does not annihilate and its particle number is conserved~\cite{Zurek:2013wia,Petraki:2013wwa}. Inside the star, we neglect any nongravitational interaction between DM and ordinary matter. The DM thus forms a separate fluid that coexists with the quark matter, exerts its own pressure, and couples to the quark fluid only through the common gravitational field~\cite{Sandin:2008db,Leung:2011zz,Ellis:2018bkr,Mukhopadhyay:2015xhs,Lopes:2023uxi}. For the same reason, the dark fluid does not exchange heat with the quark matter and is not heated by the diffusing neutrinos. We therefore treat the DM as cold, $T_D = 0$, in all three evolutionary stages, as in Ref.~\cite{Issifu:2025jac}, so that its EOS is the same in all stages and the stage dependence enters only through the quark EOS. In addition to gravity, the DM particles may interact among themselves. We describe this self-interaction by the exchange of a massive vector mediator in the mean-field approximation, in the same way as the vector interaction among quarks in Sec.~\ref{subsec:vector}.
% ---------------------------

At zero temperature the DM particles fill all momentum states up to the Fermi momentum $k_F$, and
the number density, kinetic energy density, and kinetic pressure are
\begin{align}
n_\chi &= \frac{g_\chi k_F^3}{6\pi^2} ,
\label{eq:n_chi}\\
\varepsilon_{\rm kin} &= \frac{g_\chi}{2\pi^2}\int_0^{k_F} p^2 \sqrt{p^2 + m_\chi^2}\; dp ,
\label{eq:e_chi}\\
P_{\rm kin} &= \frac{g_\chi}{6\pi^2}\int_0^{k_F} \frac{p^4}{\sqrt{p^2 + m_\chi^2}}\; dp ,
\label{eq:P_chi}
\end{align}
with spin degeneracy $g_\chi = 2$. The self-interaction is mediated by a vector field of mass $m_I$ that couples to the DM particles with coupling $g_I$. As for the quarks, in the mean-field approximation it contributes equally to the energy density and the pressure, so that the DM EOS reads~\cite{Narain:2006kx}
\begin{gather}
\varepsilon_D = \varepsilon_{\rm kin} + \frac{1}{2}\, G_\chi\, n_\chi^2 , \quad
P_D = P_{\rm kin} + \frac{1}{2}\, G_\chi\, n_\chi^2 , \nonumber\\
G_\chi = \left(\frac{g_I}{m_I}\right)^2 ,
\label{eq:DM_EOS}
\end{gather}
and the chemical potential is $\mu_\chi = (k_F^2 + m_\chi^2)^{1/2} + G_\chi n_\chi$, which
satisfies $\varepsilon_D + P_D = \mu_\chi n_\chi$. The squared sound speed follows as
\begin{equation}
c_{s,D}^2 = \frac{dP_D}{d\varepsilon_D}
= \frac{1}{\mu_\chi}\left[ \frac{k_F^2}{3\,(k_F^2 + m_\chi^2)^{1/2}} + G_\chi n_\chi \right] .
\label{eq:cs_chi}
\end{equation}
For free DM ($G_\chi = 0$), the EOS interpolates between the nonrelativistic limit, $P_D \propto n_\chi^{5/3}$ for $k_F \ll m_\chi$, and the ultrarelativistic limit, $P_D \to \varepsilon_D/3$ for $k_F \gg m_\chi$. The repulsive self-interaction grows as $n_\chi^2$ and dominates at high density, where $P_D \to \varepsilon_D$ and $c_{s,D}^2 \to 1$ from below, so that the EOS remains causal.

% -------------------------

We consider DM particle masses $m_\chi = 0.3$, $1$, and $3$~GeV and self-interaction strengths $G_\chi = 0$, $0.05$, and $0.20$~fm$^2$, given in the same units as $G_V$. For a coupling $g_I = 1$, the nonzero values correspond to mediator masses $m_I \simeq 0.88$ and $0.44$~GeV, comparable to the DM mass, and the strength of the self-interaction is better characterized by the dimensionless ratio $\xi = m_\chi/m_I = m_\chi G_\chi^{1/2}$, since the ratio of the interaction to the kinetic pressure grows as $\xi^2 = G_\chi m_\chi^2$; $\xi$ corresponds to the parameter $y = \xi/\sqrt{2}$ of Ref.~\cite{Narain:2006kx}. Our parameter set covers $\xi$ from about $0.3$ for $m_\chi = 0.3$~GeV to about $7$ for $m_\chi = 3$~GeV and thus spans the range from weak to strong self-interaction. These values are compatible with astrophysical bounds on DM self-scattering. In the Born approximation and for a heavy mediator~\cite{Tulin:2017ara}, the self-interaction cross section is $\sigma_{\chi\chi} \simeq G_\chi^2 m_\chi^2/4\pi$, which gives $\sigma_{\chi\chi}/m_\chi \simeq 1.4 \times 10^{-3}$~cm$^2$\,g$^{-1}$ for our strongest case ($m_\chi = 3$~GeV, $G_\chi = 0.20$~fm$^2$). This is about three orders of magnitude below the limit $\sigma_{\chi\chi}/m_\chi \lesssim 1$~cm$^2$\,g$^{-1}$ inferred from merging galaxy clusters~\cite{Markevitch:2003at,Randall:2008ppe,Tulin:2017ara}. For this case the Born parameter is $\alpha_\chi m_\chi/m_I \simeq 0.5$, with $\alpha_\chi = g_I^2/4\pi$, so that this estimate is only approximate. For a repulsive interaction, however, the scattering length is at most of the order of the range $1/m_I$, so that $\sigma_{\chi\chi} \lesssim 4\pi/m_I^2$, which gives $\sigma_{\chi\chi}/m_\chi \lesssim 5\times10^{-3}$~cm$^2$\,g$^{-1}$, still well below this limit.
% ------------------------------------- 
% FIGURE - 03
\begin{figure}[tbp]
\centering
\includegraphics[width=\columnwidth]{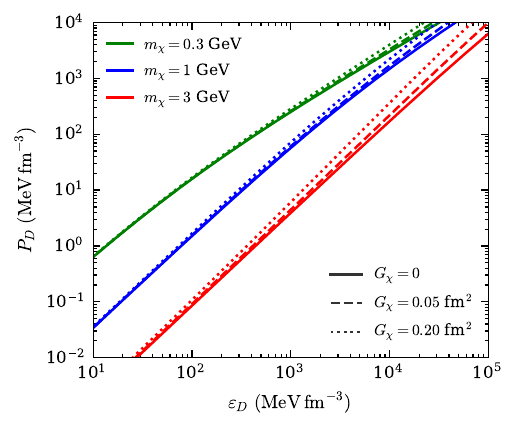}
\caption{Pressure versus energy density of cold fermionic DM for $m_\chi = 0.3$, $1$, and $3$~GeV (colors) and $G_\chi = 0$, $0.05$, and $0.20$~fm$^2$ (line styles). The EOS is the same in all three evolutionary stages.}
\label{fig:dm_eos}
\end{figure}
% ---------------------------------------
\section{Two-fluid stellar structure}
\label{sec:structure}
% -------------------------------
\subsection{Structure equations}
\label{subsec:twofluid}
Since quark matter and DM interact only through gravity, the energy-momentum tensor of the star is the sum of two perfect-fluid tensors, $T^{\mu\nu} = T^{\mu\nu}_Q + T^{\mu\nu}_D$, each of which is conserved separately, $\nabla_\mu T^{\mu\nu}_Q = \nabla_\mu T^{\mu\nu}_D = 0$. For a static, spherically symmetric star, the Einstein equations then lead to the two-fluid Tolman--Oppenheimer--Volkoff (TOV) equations~\cite{PhysRev.55.364,PhysRev.55.374,Sandin:2008db,Ciarcelluti:2010ji,Leung:2011zz}
\begin{align}
\frac{dP_j}{dr} &= -\,\frac{\left(\varepsilon_j + P_j\right)
\left[ m(r) + 4\pi r^3 P(r) \right]}{r\left[ r - 2m(r) \right]} ,
\label{eq:TOV_P}\\
\frac{dm_j}{dr} &= 4\pi r^2 \varepsilon_j , \qquad j = Q, D ,
\label{eq:TOV_m}
\end{align}
where $P = P_Q + P_D$ is the total pressure and $m = m_Q + m_D$ the total gravitational mass enclosed within radius $r$; here and in the following we use $G = c = 1$ in the structure equations. Each fluid is held up against gravity only by the gradient of its own pressure, while the gravitational field that it feels, through $m(r)$ and $P(r)$, is produced by both fluids together. This is the only coupling between the two components. The system is closed by the EOS of each fluid: for the quark matter, the EOS of the evolutionary stage under consideration, and for the DM, the stage-independent EOS of Eq.~(\ref{eq:DM_EOS}).
% -------------------------

The equations are integrated outward from the center, where $m_Q(0) = m_D(0) = 0$ and the pressures take the central values $P_{Q,c}$ and $P_{D,c}$. Each fluid ends at its own radius $R_j$, defined by $P_j(R_j) = 0$. Because quark matter is self-bound, $P_Q$ vanishes at a finite energy density, and the quark component has a sharp surface at $R_Q$, whereas the DM pressure decreases smoothly to zero. Beyond the smaller of the two radii only the other fluid is integrated, while the fluid that has ended contributes its total mass to $m(r)$. Two configurations can arise: a DM core, with $R_D < R_Q$, in which the DM is confined to the interior of the quark star, and a DM halo, with $R_D > R_Q$, in which the DM extends beyond the quark surface. With $M_Q = m_Q(R_Q)$ and $M_D = m_D(R_D)$, the total gravitational mass of the star, its outer radius, and the DM mass fraction are
\begin{equation}
M = M_Q + M_D , \quad R = \max(R_Q, R_D) , \quad F_D = \frac{M_D}{M} .
\label{eq:MR}
\end{equation}
Since the DM does not emit or absorb electromagnetic radiation, the visible surface of the star is at $R_Q$ also in the halo case, whereas the gravitational mass measured at large distance is the total mass $M$.
% ---------------------------------

To follow a given star through the evolutionary stages, we also compute the numbers of baryons and of DM particles,
\begin{gather}
N_B = \int_0^{R_Q} \frac{4\pi r^2\, n_B(r)}{\sqrt{1 - 2m(r)/r}}\; dr , \nonumber\\
N_\chi = \int_0^{R_D} \frac{4\pi r^2\, n_\chi(r)}{\sqrt{1 - 2m(r)/r}}\; dr ,
\label{eq:NB}
\end{gather}
where the factor $[1 - 2m(r)/r]^{-1/2}$ accounts for the proper volume of each shell, and express them through the baryon mass $M_B = m_n N_B$, with $m_n$ the neutron mass, and the DM rest mass $M_\chi = m_\chi N_\chi$. During the evolution the star loses energy and lepton number through the emitted neutrinos, so that its gravitational mass decreases. The number of baryons, however, is conserved, and so is the number of DM particles, since the DM is asymmetric and interacts with the quark matter only through gravity; we neglect the capture of additional DM on the short timescales considered here. An evolutionary sequence is hence described by stars with the same $N_B$ and $N_\chi$ in the three stages, rather than by stars with the same $M$ or $F_D$.
% ----------------------------

We use two kinds of sequences. To compare stars with the same DM content, we construct sequences at fixed $F_D = 1$, $3$, and $5\%$, with additional values between $3$ and $5\%$ for $m_\chi = 3$~GeV, together with pure quark stars. To follow individual stars, we construct sequences at fixed $N_\chi$. Along these sequences, the largest baryon mass $M_{B,\rm max}$ is the limit of stability at fixed $N_B$ and $N_\chi$. At this point the variations of $N_B$ and $N_\chi$ with the central pressures are parallel, i.e., the Jacobian of $(N_B, N_\chi)$ with respect to $(P_{Q,c}, P_{D,c})$ vanishes,
\begin{equation}
\frac{\partial N_B}{\partial P_{Q,c}}\frac{\partial N_\chi}{\partial P_{D,c}}
-\frac{\partial N_B}{\partial P_{D,c}}\frac{\partial N_\chi}{\partial P_{Q,c}} = 0 ,
\label{eq:stability}
\end{equation}
which is the particle-number criterion for two-component stars~\cite{Henriques:1989ez,Chen:2025soc}, and it marks the onset of instability of the fundamental radial mode~\cite{Caballero:2024qtv,Chen:2025soc}, which we do not compute explicitly. Since neutrino diffusion is far slower than the radial oscillations, the perturbations keep $s_Q/n_B$ and $Y_l$ fixed, consistent with our stages. The maximum of a fixed-$F_D$ sequence does not in general coincide with this limit~\cite{Kain:2020zjs,Caballero:2024qtv}; in our calculations, however, the gravitational mass at $M_{B,\rm max}$ agrees with the maximum mass of the corresponding fixed-$F_D$ sequence to within $0.004\,M_\odot$, except in the three cases in which the DM core reaches its own mass limit (Sec.~\ref{subsec:res_cold}). 
% ---------------------------------
\subsection{Tidal deformability}
\label{subsec:tidal}
In a binary system the tidal field $\mathcal{E}_{ij}$ of the companion induces a quadrupole moment $Q_{ij} = -\lambda\, \mathcal{E}_{ij}$ in each star, and the tidal deformability $\lambda$ affects the phase of the emitted gravitational waves~\cite{Flanagan:2007ix,Hinderer:2007mb}. It is usually expressed through the dimensionless quantity
\begin{equation}
\Lambda = \frac{\lambda}{M^5} = \frac{2}{3}\, \frac{k_2}{C^5} , \qquad C = \frac{M}{R} ,
\label{eq:Lambda}
\end{equation}
where $k_2$ is the quadrupolar tidal Love number. To compute $k_2$ we solve the equation for the static, even-parity $l = 2$ perturbation of the metric. Since the two fluids interact only through gravity, each of them responds to the tidal field according to its own EOS, and the perturbation equation keeps its single-fluid form with the total energy density and pressure, except for the term that contains the compressibility of the matter, which becomes a sum over the two fluids~\cite{Ellis:2018bkr,Leung:2022wcf}. For $y(r) = r H'(r)/H(r)$, where $H$ is the metric perturbation, it reads
\begin{equation}
r\,\frac{dy}{dr} = -\,y^2 - y\, e^{\beta}\left[ 1 + 4\pi r^2 \left( P - \varepsilon \right) \right]
- r^2 \mathcal{Q}(r) ,
\label{eq:tidal_y}
\end{equation}
with
\begin{equation}
\mathcal{Q} = 4\pi e^{\beta}\left[ 5\varepsilon + 9P
+ \sum_{j = Q, D} \left( \varepsilon_j + P_j \right) \frac{d\varepsilon_j}{dP_j} \right]
- \frac{6\, e^{\beta}}{r^2} - \nu'^2 ,
\label{eq:tidal_Q}
\end{equation}
$e^{\beta} = [1 - 2m(r)/r]^{-1}$, and $\nu' = 2[m(r) + 4\pi r^3 P(r)]/\{r[r - 2m(r)]\}$. In the hot stages, $d\varepsilon_Q/dP_Q$ is taken along the isentrope of the stage, at fixed $s_Q/n_B$ and, in the neutrino-trapped stage, fixed $Y_l$. Equation~(\ref{eq:tidal_y}) is integrated together with Eqs.~(\ref{eq:TOV_P}) and (\ref{eq:TOV_m}), starting from $y(0) = 2$. The self-bound quark matter has a finite energy density $\varepsilon_s$ at its surface, and the discontinuity of $\varepsilon_Q$ at $R_Q$ requires the matching condition~\cite{Damour:2009vw,Postnikov:2010yn,Takatsy:2020bnx}
\begin{equation}
y(R_Q^+) = y(R_Q^-) - \frac{4\pi R_Q^3\, \varepsilon_s}{m(R_Q) + 4\pi R_Q^3 P(R_Q)},
\label{eq:tidal_jump}
\end{equation}
where $P(R_Q) = P_D(R_Q)$ is nonzero only when the DM forms a halo. The DM density goes continuously to zero at $R_D$ and needs no such correction. With $y_R = y(R)$ at the outer radius $R = \max(R_Q, R_D)$ and $C = M/R$, the Love number is
\begin{align}
k_2 &= \frac{8}{5}\, \frac{C^5 (1 - 2C)^2 \left[ 2 + 2C (y_R - 1) - y_R \right]}{\mathcal{D}} ,
\label{eq:k2}\\
\mathcal{D} &= 2C \left[ 6 - 3y_R + 3C (5 y_R - 8) \right] \nonumber\\
&\quad + 4C^3 \left[ 13 - 11 y_R + C (3 y_R - 2) + 2 C^2 (1 + y_R) \right] \nonumber\\
&\quad + 3 (1 - 2C)^2 \left[ 2 - y_R + 2C (y_R - 1) \right] \ln (1 - 2C) .
\nonumber
\end{align}
For a DM halo the outer radius is $R_D$, so that the tidal response is that of the whole two-fluid configuration.
% ---------------------------------------
\section{Results}
\label{sec:results}
% ------------------------------------
% FIGURE - 04
\begin{figure*}[tbp]
\centering
\includegraphics[width=\textwidth]{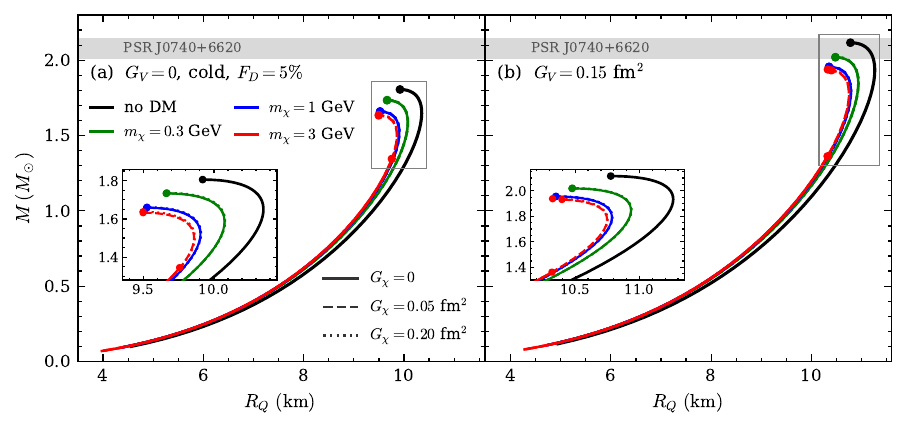}
\caption{Mass--radius relations of cold strange stars with a DM mass fraction $F_D = 5\%$ for (a)~$G_V = 0$ and (b)~$G_V = 0.15$~fm$^2$. Colors denote the DM particle mass, $m_\chi = 0.3$ (green), $1$ (blue), and $3$~GeV (red), and line styles the self-interaction, $G_\chi = 0$ (solid), $0.05$ (dashed), and $0.20$~fm$^2$ (dotted); the black curves are the stars without DM. Only the stable branches are shown, ending at the maximum masses (dots), and $R_Q$ is the radius of the quark surface. For $m_\chi = 3$~GeV and $G_\chi = 0$ the sequences end where the DM core reaches its own mass limit (Sec.~\ref{subsec:res_cold}). The insets magnify the regions of the maximum masses. The gray bands show the mass of PSR~J0740$+$6620, $2.08 \pm 0.07\,M_\odot$~\cite{Fonseca:2021wxt}.}
\label{fig:MR}
\end{figure*}
% ----------------------
% FIGURE - 05
\begin{figure*}[t]
\centering
\includegraphics[width=\textwidth]{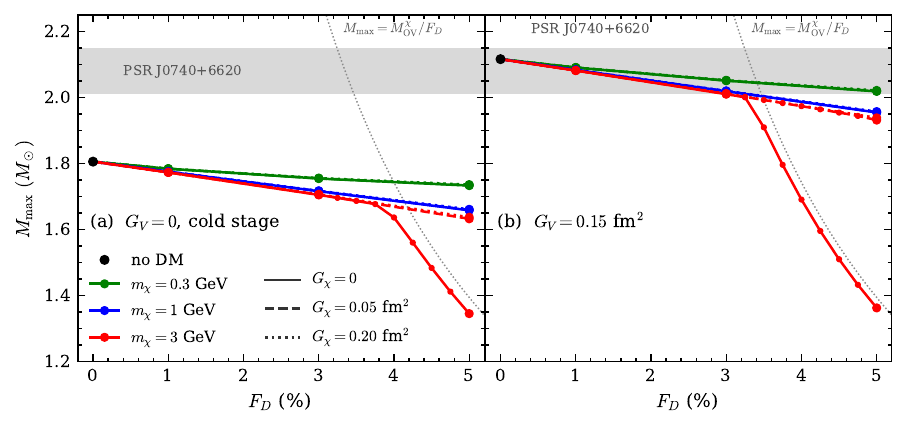}
\caption{Maximum mass of cold DM-admixed strange stars as a function of the DM mass fraction $F_D$, for (a)~$G_V = 0$ and (b)~$G_V = 0.15$~fm$^2$. Colors, line styles, and gray bands as in Fig.~\ref{fig:MR}. Large dots are the fractions $F_D = 0$, $1$, $3$, and $5\%$; small dots are additional fractions for $m_\chi = 3$~GeV with $G_\chi = 0$ and, in panel (b), $0.05$~fm$^2$. The thin dotted lines are $M_{\rm max} = M_{\rm OV}^\chi/F_D$, where $M_{\rm OV}^\chi = 0.070\,M_\odot$ is the maximum mass of a star made of free DM with $m_\chi = 3$~GeV.}
\label{fig:Mmax_FD}
\end{figure*}
% --------------------
% FIGURE - 06
\begin{figure*}[t]
\centering
\includegraphics[width=\textwidth]{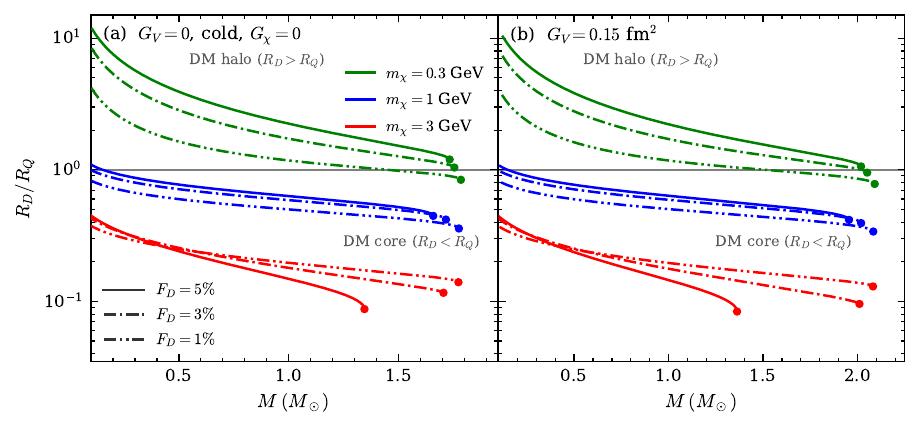}
\caption{Ratio of the DM radius to the quark radius, $R_D/R_Q$, along the stable branches as a function of the total mass, for cold stars with $G_\chi = 0$ and (a)~$G_V = 0$ and (b)~$G_V = 0.15$~fm$^2$. Colors denote $m_\chi$ as in Fig.~\ref{fig:MR}, and line styles the DM fraction, $F_D = 5\%$ (solid), $3\%$ (dash-dotted), and $1\%$ (dash-dot-dotted). Dots mark the maximum-mass stars. A ratio $R_D/R_Q > 1$ corresponds to a DM halo and $R_D/R_Q < 1$ to a DM core.}
\label{fig:core_halo}
\end{figure*}
%\vspace{4mm}
% FIGURE - 07
\begin{figure*}[tbp]
\centering
\includegraphics[width=\textwidth]{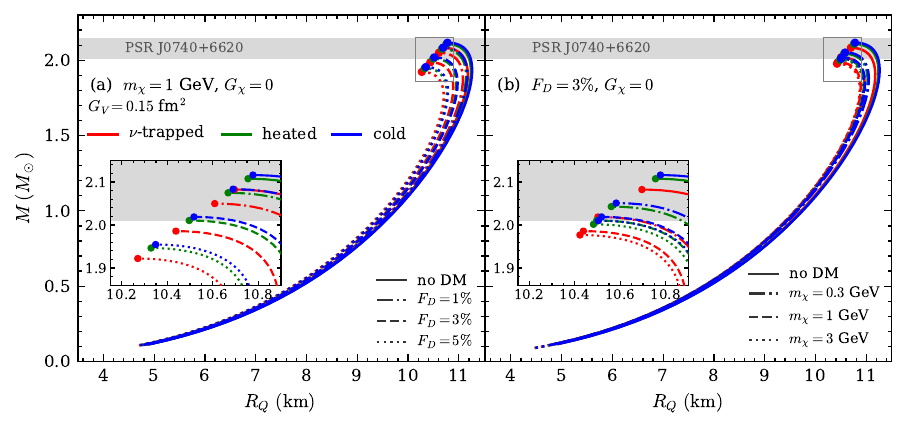}
\caption{Mass--radius relations of DM-admixed strange stars in the neutrino-trapped (red), heated (green), and cold (blue) stages for $G_V = 0.15$~fm$^2$ and $G_\chi = 0$; here colors denote the stage. (a)~$m_\chi = 1$~GeV with $F_D = 0$ (solid), $1\%$ (dash-dotted), $3\%$ (dashed), and $5\%$ (dotted). (b)~Without DM (solid) and with $F_D = 3\%$ for $m_\chi = 0.3$ (dash-dotted), $1$ (dashed), and $3$~GeV (dotted); the dashed curves in the two panels are the same stars. Only the stable branches are shown, ending at the maximum masses (dots). The insets magnify the regions of the maximum masses. Gray bands as in Fig.~\ref{fig:MR}.}
\label{fig:MR_stages}
\end{figure*}
% --------------------------------
% FIGURE - 08 
\begin{figure*}[tbp]
\centering
\includegraphics[width=\textwidth]{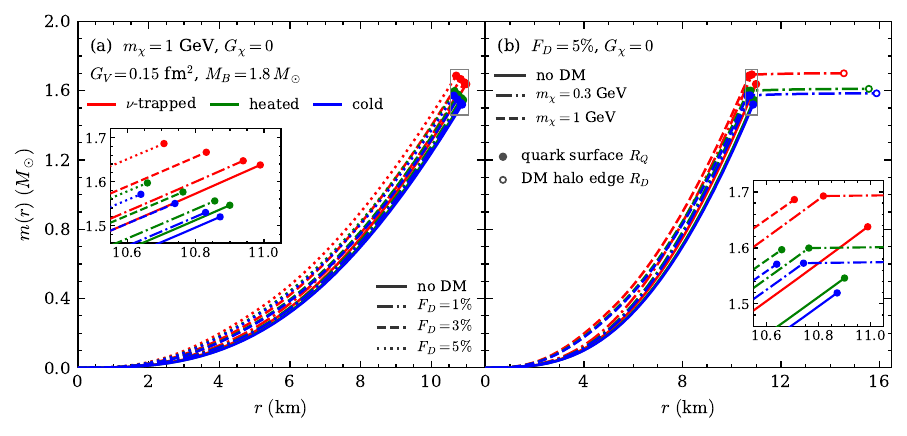}
\caption{Enclosed gravitational mass $m(r) = m_Q(r) + m_D(r)$ inside a star with baryon mass $M_B = 1.8\,M_\odot$, followed through the neutrino-trapped (red), heated (green), and cold (blue) stages, for $G_V = 0.15$~fm$^2$ and $G_\chi = 0$. With DM, the DM rest mass is also fixed, such that the cold star has the stated $F_D$. (a)~$m_\chi = 1$~GeV with $F_D = 0$, $1$, $3$, and $5\%$; (b)~without DM and with $F_D = 5\%$ for $m_\chi = 0.3$ and $1$~GeV (line styles as in Fig.~\ref{fig:MR_stages}); for $m_\chi = 3$~GeV no such star exists, because its DM core would exceed the limit $M_{D,\rm lim}$ (Sec.~\ref{subsec:res_cold}). Filled dots mark the quark surface $R_Q$ and open circles the edge $R_D$ of the DM halo. The insets magnify the region near the quark surface.}
\label{fig:m_r}
\end{figure*}
% --------------------------
% ----------------------- 
\subsection{Equation of state of quark matter}
\label{subsec:res_eos} 
Figure~\ref{fig:P_eps} shows the EOS of quark matter in the three stages. In all cases the pressure vanishes at a finite energy density, $\varepsilon_0 \simeq 250$--$275$~MeV\,fm$^{-3}$, which reflects the self-bound nature of SQM. For massless quarks and $G_V = 0$ this point would lie at $\varepsilon_0 = 4B$; the finite strange-quark mass shifts it to higher values, and the vector interaction to lower ones. For $G_V = 0$ the EOS is nearly linear, $P_Q \simeq (\varepsilon_Q - \varepsilon_0)/3$, with a squared sound speed close to the conformal value $1/3$. The vector interaction stiffens the EOS considerably. Since its contribution $P_V = \varepsilon_V \propto n_q^2$ [Eq.~(\ref{eq:vector})] grows faster with density than the kinetic pressure of the quarks, its importance increases toward the center, and for $G_V = 0.15$~fm$^2$ the squared sound speed exceeds $0.5$ at the highest densities reached in the stars.
% -------------------------

The thermal and neutrino contributions have a much smaller effect. Since the temperature remains far below the quark chemical potentials ($T/\mu \lesssim 0.1$), the three stages are almost indistinguishable for $G_V = 0$. The differences are visible only near the surface, where the total pressure is small (inset of Fig.~\ref{fig:P_eps}): the additional pressure of the trapped neutrinos and of the thermal excitations moves the zero-pressure point to slightly lower energy density. For $G_V \neq 0$, in contrast, the neutrino-trapped EOS is noticeably softer than those of the heated and cold stages. At a given energy density, part of the energy is carried by the trapped leptons, so that the baryon density, and with it the quark density, is lower. Since the vector term depends only on the quark density, it is correspondingly smaller, and the total pressure is lower. Without the vector interaction this softening does not occur, because all constituents are then nearly ultrarelativistic, so that $P_Q \simeq (\varepsilon_Q - 4B)/3$ regardless of how the energy is shared among quarks, leptons, and gluons.
% -----------------------

The temperature along the two isentropes is shown in Fig.~\ref{fig:T_nB}. It increases with density in both stages. This behavior follows from the strong degeneracy of the quarks. For a degenerate quark gas the entropy per quark is proportional to $T/\mu$, so that at fixed $s_Q/n_B$ the temperature scales as $T \propto \mu \propto n_B^{1/3}$, which our results follow closely. If the quarks were the only carriers of entropy, doubling $s_Q/n_B$ would double the temperature. The ratio between the two stages is instead about $2.3$ at all densities, because in the neutrino-trapped stage part of the entropy is carried by the neutrinos and the additional electrons. The open circles coincide with the lines, since at a given density the temperature does not depend on $G_V$ (Sec.~\ref{subsec:equilibrium}); the vector interaction only moves the surface to lower density, where the temperature is slightly lower. The thermal gluons lower the temperature by at most about $1\%$. The central temperatures of the maximum-mass stars are listed in Table~\ref{tab:pureQS}. 
% ------------------------------------
% TABLE - 03
\begin{table}[t]
\caption{\label{tab:pureQS}Maximum-mass configurations of strange quark stars without DM in the
three evolutionary stages, for $B^{1/4} = 150$~MeV and the four values of $G_V$: central baryon
density $n_{B,c}$ in units of $n_0 = 0.16$~fm$^{-3}$, central temperature $T_c$, baryon mass
$M_{B,\rm max}$, maximum gravitational mass $M_{\rm max}$, and radius $R$. Without DM the
maximum of the baryon mass coincides with that of the gravitational mass.}
\setlength{\tabcolsep}{3.5pt}
\begin{ruledtabular}
\begin{tabular}{llccccc}
$G_V$ & Stage & $n_{B,c}$ & $T_c$ & $M_{B,\rm max}$ & $M_{\rm max}$ & $R$ \\
(fm$^2$) & & ($n_0$) & (MeV) & ($M_\odot$) & ($M_\odot$) & (km) \\
\hline
\multirow{3}{*}{0.00} & $\nu$-trapped & 6.97 & 12.9 & 2.126 & 1.816 & 9.97 \\
 & heated & 7.55 & 30.1 & 2.272 & 1.808 & 9.93 \\
 & cold & 7.73 & 0 & 2.318 & 1.806 & 9.92 \\
\hline
\multirow{3}{*}{0.05} & $\nu$-trapped & 6.53 & 12.6 & 2.232 & 1.919 & 10.24 \\
 & heated & 7.00 & 29.3 & 2.402 & 1.925 & 10.24 \\
 & cold & 7.13 & 0 & 2.457 & 1.928 & 10.24 \\
\hline
\multirow{3}{*}{0.10} & $\nu$-trapped & 6.14 & 12.4 & 2.318 & 2.006 & 10.48 \\
 & heated & 6.53 & 28.7 & 2.505 & 2.023 & 10.51 \\
 & cold & 6.64 & 0 & 2.565 & 2.029 & 10.53 \\
\hline
\multirow{3}{*}{0.15} & $\nu$-trapped & 5.82 & 12.1 & 2.390 & 2.082 & 10.70 \\
 & heated & 6.14 & 28.1 & 2.588 & 2.108 & 10.75 \\
 & cold & 6.23 & 0 & 2.651 & 2.116 & 10.78 \\
\end{tabular}
\end{ruledtabular}
\end{table}
% ------------------------------
\subsection{Equation of state of dark matter}
\label{subsec:res_dm}
The DM EOS is shown in Fig.~\ref{fig:dm_eos}. At a given energy density, lighter DM particles are more numerous, $n_\chi \simeq \varepsilon_D/m_\chi$, and so have a larger Fermi momentum and a higher degeneracy pressure. In the nonrelativistic regime this gives $P_D \propto \varepsilon_D^{5/3}\, m_\chi^{-8/3}$, so that light DM is much stiffer than heavy DM. The transition to the relativistic regime, $k_F = m_\chi$, occurs at an energy density that scales as $m_\chi^4$; it lies within the range of Fig.~\ref{fig:dm_eos} for $m_\chi = 0.3$ and $1$~GeV, but well above it for $m_\chi = 3$~GeV. The self-interaction leaves the EOS unchanged at low density, where $G_\chi n_\chi^2/2$ is negligible, and stiffens it at high density. Its effect grows with the DM mass, in line with $\xi^2 = G_\chi m_\chi^2$ (Sec.~\ref{subsec:dm}): at $\varepsilon_D = 10^4$~MeV\,fm$^{-3}$, $G_\chi = 0.20$~fm$^2$ raises the pressure by about $30\%$ for $m_\chi = 0.3$~GeV but by a factor of $2$ for $m_\chi = 3$~GeV.
% --------------------------

A comparison with the quark-matter EOS of Fig.~\ref{fig:P_eps} anticipates the structure of the admixed stars. At $\varepsilon = 10^3$~MeV\,fm$^{-3}$, a typical energy density inside the quark star, DM with $m_\chi = 0.3$~GeV has a pressure comparable to that of cold SQM with $G_V = 0.15$~fm$^2$, whereas the pressure of DM with $m_\chi = 1$ and $3$~GeV is lower by factors of about $5$ and $50$--$80$, respectively. Light DM can therefore support itself against gravity over a large volume and can extend beyond the quark surface, whereas heavy DM is too soft to do so and concentrates in a compact core at the center of the star.
% ---------------------------

A useful reference for the admixed stars is provided by self-gravitating configurations made of DM alone. For free DM their maximum mass is the Oppenheimer--Volkoff limit~\cite{PhysRev.55.374,Narain:2006kx}, $M_{\rm max} = 0.384\, M_{\rm Pl}^3/m_\chi^2$, where $M_{\rm Pl}$ is the Planck mass. It falls from about $7\,M_\odot$ for $m_\chi = 0.3$~GeV to $0.63\,M_\odot$ for $1$~GeV and $0.070\,M_\odot$ for $3$~GeV. The self-interaction raises the maximum mass, and its effect grows with $m_\chi$: for $G_\chi = 0.20$~fm$^2$ the increase is only $2.5\%$ for $m_\chi = 0.3$~GeV but $143\%$ for $m_\chi = 3$~GeV. The small maximum mass of heavy free DM is directly relevant for the admixed stars. For $m_\chi = 3$~GeV and $G_\chi = 0$, a DM mass fraction of $5\%$ in a star of $1.3$--$1.4\,M_\odot$ corresponds to $M_D \approx 0.07\,M_\odot$, close to this limit. In this case the stability of the star can be limited by the DM component itself rather than by the quark matter. For larger fractions no stable configurations with heavy free DM exist in this mass range, which is one reason why we restrict the DM mass fraction to $F_D \le 5\%$.
% ---------------------------
\subsection{Cold DM-admixed stars}
\label{subsec:res_cold}
% ---------------------

Cold DM-admixed strange stars have been studied before~\cite{Mukhopadhyay:2015xhs,Panotopoulos:2017pgv,PhysRevD.104.083016,Lopes:2023uxi,Marzola:2024ame,Yang:2024vjx}. We summarize their properties in our model, which serve as the reference for the evolution discussed in Secs.~\ref{subsec:res_stages} and \ref{subsec:res_evolution}. Figure~\ref{fig:MR} shows the mass--radius relations for $F_D = 5\%$, the largest DM fraction considered, for the softest and the stiffest quark matter. As discussed in Sec.~\ref{subsec:twofluid}, we give the radius $R_Q$ of the visible quark surface and the total gravitational mass $M$. The DM lowers the maximum mass and makes the star more compact, and its effect grows with the DM particle mass: for $G_V = 0.15$~fm$^2$, the maximum mass of the star without DM (Table~\ref{tab:pureQS}) is reduced by about $5\%$, $8\%$, and $9\%$ for $m_\chi = 0.3$, $1$, and $3$~GeV (the last for $G_\chi > 0$), and the radius of a $1.4\,M_\odot$ star by $0.2$--$0.3$~km. Heavy DM is more effective because it is concentrated in a compact core, where it adds to the gravitational pull on the quark matter without contributing to its pressure support, whereas light DM is spread over a much larger volume and can extend beyond $R_Q$. The self-interaction has almost no effect for $m_\chi = 0.3$ and $1$~GeV, since the interaction term is small at the DM densities reached in these stars (Sec.~\ref{subsec:res_dm}).
% ---------------------------------------
 
The dependence on the DM fraction is shown in Fig.~\ref{fig:Mmax_FD}. For $m_\chi = 1$ and $3$~GeV the maximum mass decreases almost linearly with $F_D$, by $0.03$--$0.04\,M_\odot$ per percent of DM, for $m_\chi = 3$~GeV and $G_\chi = 0$ up to the break discussed below. For $m_\chi = 0.3$~GeV the decrease is weaker and becomes smaller as $F_D$ grows, because a growing part of the light DM resides beyond the quark surface. Even in this case, in which the DM forms a halo, the DM does not increase the maximum mass of the star for the fractions considered here, $F_D \le 5\%$.
% ----------------------------

The curve for $m_\chi = 3$~GeV and $G_\chi = 0$ shows a break at $F_D \simeq 3$--$4\%$, beyond which the maximum mass falls steeply and closely follows the line $M_{\rm max} = M_{\rm OV}^\chi/F_D$. Beyond the break, the DM mass of the maximum-mass star saturates at $M_{D,\rm lim} \simeq 0.067\,M_\odot$. This is the largest mass that the DM core can have inside the quark matter, slightly below the limit of an isolated DM star (Sec.~\ref{subsec:res_dm}), because the gravity of the surrounding quark matter compresses the core. The maximum mass is then set by the DM, $M_{\rm max} \simeq M_{D,\rm lim}/F_D$, rather than by the quark matter, and it is nearly independent of the quark EOS, about $1.35\,M_\odot$ at $F_D = 5\%$ for both values of $G_V$. Since the break occurs where $F_D M_{\rm max}$ reaches $M_{D,\rm lim}$, it appears at a lower fraction for the stiffer quark matter, whose stars are heavier. A repulsive self-interaction raises $M_{D,\rm lim}$ and removes the break within $F_D \le 5\%$; only for $G_\chi = 0.05$~fm$^2$ and $G_V = 0.15$~fm$^2$ does the DM core come close to its limit at $F_D = 5\%$, and at fixed DM rest mass it reaches this limit (Sec.~\ref{subsec:res_evolution}).
% ----------------------

The mass of PSR~J0740$+$6620~\cite{Fonseca:2021wxt} constrains these configurations. For $G_V = 0$ none of the stars reaches the observed mass, with or without DM. For $G_V = 0.15$~fm$^2$ the maximum mass without DM lies within the $1\sigma$ band, and it remains there for DM fractions up to $3\%$ for all three DM masses (at the lower edge of the band for $m_\chi = 3$~GeV), but at $F_D = 5\%$ only for $m_\chi = 0.3$~GeV. Within our model, a stiff quark EOS is therefore required for DM-admixed strange stars to be compatible with the heaviest known pulsars, and the DM content of such a star cannot exceed a few percent unless the DM is light.
% ----------------------------------
% TABLE - 04
\begin{table}[t]
\caption{\label{tab:DMstages}Strange quark stars with and without DM in the three evolutionary stages, for $G_V = 0.15$~fm$^2$ and $G_\chi = 0$; the DM cases are labeled by $m_\chi$ and $F_D$. $M_{B,\rm max}$: largest baryon mass at fixed DM rest mass (Fig.~\ref{fig:M_MB}); $M_{1.8}$: gravitational mass of the star with $M_B = 1.8\,M_\odot$ (Fig.~\ref{fig:M_MB}). In these two columns the DM rest mass is that of the star with $M_B = 1.8\,M_\odot$, and $F_D$ refers to its cold stage. $M_{\rm max}$ and $R_Q$: maximum mass and quark radius on the sequence at fixed $F_D$ (Fig.~\ref{fig:MR_stages}); $\Lambda_{1.4}$: tidal deformability at $M = 1.4\,M_\odot$ on the same sequence (Fig.~\ref{fig:Lambda}). The rows without DM repeat the entries of Table~\ref{tab:pureQS} for $G_V = 0.15$~fm$^2$ as the reference.}
\setlength{\tabcolsep}{3pt}
\begin{ruledtabular}
\begin{tabular}{llccccc}
$m_\chi$, $F_D$ & Stage & $M_{B,\rm max}$ & $M_{1.8}$ & $M_{\rm max}$ & $R_Q$ & $\Lambda_{1.4}$ \\
(GeV) & & ($M_\odot$) & ($M_\odot$) & ($M_\odot$) & (km) & \\
\hline
\multirow{3}{*}{no DM} & $\nu$-trapped & 2.390 & 1.638 & 2.082 & 10.70 & 392 \\
 & heated & 2.588 & 1.546 & 2.108 & 10.75 & 400 \\
 & cold & 2.651 & 1.520 & 2.116 & 10.78 & 404 \\
\hline
\multirow{3}{*}{1, 1\%} & $\nu$-trapped & 2.345 & 1.647 & 2.050 & 10.61 & 368 \\
 & heated & 2.541 & 1.556 & 2.075 & 10.67 & 376 \\
 & cold & 2.603 & 1.530 & 2.083 & 10.69 & 379 \\
\hline
\multirow{3}{*}{1, 3\%} & $\nu$-trapped & 2.253 & 1.667 & 1.986 & 10.44 & 326 \\
 & heated & 2.443 & 1.576 & 2.011 & 10.50 & 334 \\
 & cold & 2.503 & 1.550 & 2.019 & 10.52 & 337 \\
\hline
\multirow{3}{*}{1, 5\%} & $\nu$-trapped & 2.155 & 1.686 & 1.922 & 10.27 & 288 \\
 & heated & 2.338 & 1.597 & 1.947 & 10.33 & 295 \\
 & cold & 2.397 & 1.571 & 1.955 & 10.35 & 298 \\
\hline
\multirow{3}{*}{0.3, 3\%} & $\nu$-trapped & 2.274 & 1.673 & 2.019 & 10.50 & 359 \\
 & heated & 2.463 & 1.583 & 2.043 & 10.56 & 367 \\
 & cold & 2.524 & 1.557 & 2.051 & 10.58 & 370 \\
\hline
\multirow{3}{*}{3, 3\%} & $\nu$-trapped & 2.248 & 1.665 & 1.977 & 10.42 & 321 \\
 & heated & 2.437 & 1.575 & 2.002 & 10.48 & 328 \\
 & cold & 2.498 & 1.549 & 2.010 & 10.50 & 331 \\
\end{tabular}
\end{ruledtabular}
\end{table}
% -----------------------------------

Whether the DM forms a core inside the quark star or a halo around it is shown in Fig.~\ref{fig:core_halo}, where we plot the ratio $R_D/R_Q$ along the stable branches. Heavy DM forms a compact core well inside the quark star; only in very light stars does $1$~GeV DM extend slightly beyond the quark surface. Light DM behaves differently. For $m_\chi = 0.3$~GeV the DM extends far beyond the quark surface in light stars, up to $R_D/R_Q \gtrsim 10$, but the ratio decreases steadily with the stellar mass, because the deepening gravitational potential of the quark star pulls the DM inward. Whether the DM ends up as a core or a halo then depends on its fraction and on the quark EOS. For $G_V = 0.15$~fm$^2$ the halo turns into a core at $M \simeq 1.4\,M_\odot$ for $F_D = 1\%$ and just below the maximum mass for $F_D = 3\%$, whereas for $F_D = 5\%$ the DM forms a halo up to the maximum mass. For $G_V = 0$, the stars are lighter, and the halo survives up to the maximum mass already for $F_D = 3\%$. The self-interaction changes the DM distribution noticeably only for $m_\chi = 3$~GeV, where it enlarges the compact core. The two distributions are also reflected in the central energy density of the DM, which for $m_\chi = 0.3$~GeV lies below the surface energy density of SQM, whereas for $m_\chi = 1$~GeV it is of the same order as the central energy density of the quark matter and for $m_\chi = 3$~GeV far above it.
% ---------------------------------

\subsection{DM-admixed stars in the three stages}
\label{subsec:res_stages}
We now turn to the hot stages. Without DM, the maximum-mass configurations of the three stages are listed in Table~\ref{tab:pureQS}. The stage dependence of proto-strange quark stars has been discussed before~\cite{Dexheimer:2013eua,Lopes:2020dvs,Issifu:2023qoo}, and we summarize it here as the reference for the admixed stars. The maximum mass depends only weakly on the stage, and the sign of the dependence changes with the vector coupling. For $G_V = 0$ the neutrino-trapped star is slightly heavier than the heated and cold ones, because the trapped neutrinos and the thermal excitations add to the pressure. For $G_V \ge 0.05$~fm$^2$ the order is reversed, and the difference grows with $G_V$, because the neutrino-trapped EOS is softer (Sec.~\ref{subsec:res_eos}). The largest baryon mass $M_{B,\rm max}$, in contrast, increases from the neutrino-trapped to the heated and cold stages for every $G_V$. This confirms the assumption, made in earlier studies of proto-strange quark stars, that such stars evolve to the cold stage without forming a black hole~\cite{Issifu:2023qoo,Chen:2026nxv}. This increase determines the stability of the evolving star, as discussed in Sec.~\ref{subsec:res_evolution}.
% -------------------------------

Figure~\ref{fig:MR_stages} shows how the DM changes the mass--radius relations in the three stages for $G_V = 0.15$~fm$^2$. Unless stated otherwise, here and in the rest of this section we take $G_\chi = 0$: the self-interaction has almost no effect for $m_\chi = 0.3$ and $1$~GeV (Figs.~\ref{fig:MR} and \ref{fig:Mmax_FD}), and for $m_\chi = 3$~GeV it only raises the mass limit of the DM core, so that free DM gives the largest effect of the DM on the star. In each stage the DM lowers the maximum mass and makes the star more compact, as in the cold stage (Sec.~\ref{subsec:res_cold}), and the effect of the DM is nearly the same in the three stages. For $m_\chi = 1$~GeV and $F_D = 5\%$, the maximum mass is reduced by about $8\%$ in each stage, and the radius of a $1.4\,M_\odot$ star by $0.27$~km. For comparison, the radius of a $1.4\,M_\odot$ star changes by only $0.04$~km between the neutrino-trapped and the cold stage, with or without DM. The stage ordering of the maximum mass found without DM is preserved for both values of $G_V$: for $G_V = 0.15$~fm$^2$ the maximum mass increases from the neutrino-trapped to the cold stage in all DM cases (Table~\ref{tab:DMstages}). The effect of the DM is nearly independent of the stage because the DM is cold and its EOS is the same in all stages, so that the stage enters only through the quark EOS, whose stage dependence is small at the densities of these stars. Panel~(b) shows the ordering by DM mass found for the cold stars, with the curves for $m_\chi = 1$ and $3$~GeV nearly coinciding at $F_D = 3\%$, because once the DM forms a small central core, its precise size matters little. At this fraction the $3$~GeV core remains below its own mass limit, and the sequences end at a regular maximum in all three stages. 
% -----------------------------------------
\subsection{Evolution of a newborn star}
\label{subsec:res_evolution}
A comparison at fixed $F_D$ does not describe the evolution of a given star, since a star keeps its numbers of baryons and DM particles, not its mass or its DM fraction (Sec.~\ref{subsec:twofluid}). Unlike studies of DM-admixed proto-neutron stars, which typically keep the DM fraction fixed between stages~\cite{Issifu:2025jac,Issifu:2025gsq}, we follow a single star at fixed $N_B$ and $N_\chi$. We take a baryon mass $M_B = 1.8\,M_\odot$ in every stage, a typical value that corresponds to $M \simeq 1.5$--$1.6\,M_\odot$ in the cold stage. With DM, we also fix the DM rest mass $M_\chi$, chosen such that the cold star has the stated DM fraction exactly, so that the values of $F_D$ quoted in this section refer to the cold stage. Unless stated otherwise, the results are for $G_V = 0.15$~fm$^2$ and $G_\chi = 0$, and they are collected in Table~\ref{tab:DMstages}.
% ----------------------------

The enclosed mass of these stars is shown in Fig.~\ref{fig:m_r}. Without DM, the gravitational mass of the star decreases by $0.118\,M_\odot \simeq 2.1 \times 10^{53}$~erg from the neutrino-trapped to the cold stage (Table~\ref{tab:DMstages}). This is the energy carried away by the neutrinos, nearly $80\%$ of it during deleptonization. A star of given baryon number is thus heaviest and largest in its neutrino-trapped stage, because it still contains the thermal and lepton energy that the neutrinos remove later. The DM does not take part in this energy loss. Since its particle number is fixed and it is not heated, its gravitational mass stays nearly the same in all stages, whereas the quark component loses energy; as a result, the DM fraction of the star increases slightly during the evolution. A star with DM is heavier than the star without DM at the same $M_B$, by about two thirds of the DM mass for a DM core and about $80\%$ of it for the $0.3$~GeV halo, since its quark component is more compact and hence more strongly bound. The energy released differs from that of the star without DM by less than $3\%$ in all cases. The DM distribution, however, follows the evolution of the quark matter. As the quark component loses energy, the gravitational field that confines the DM weakens, and the DM expands: between the neutrino-trapped and the cold stage, the $1$~GeV core grows by about $4\%$ in radius for $F_D = 3\%$, and the $0.3$~GeV halo by about $9\%$ for $F_D = 5\%$, whereas the compact $3$~GeV core, which is bound most strongly, hardly changes.
% ----------------------------
% FIGURE - 09
\begin{figure*}[t]
\centering
\includegraphics[width=\textwidth]{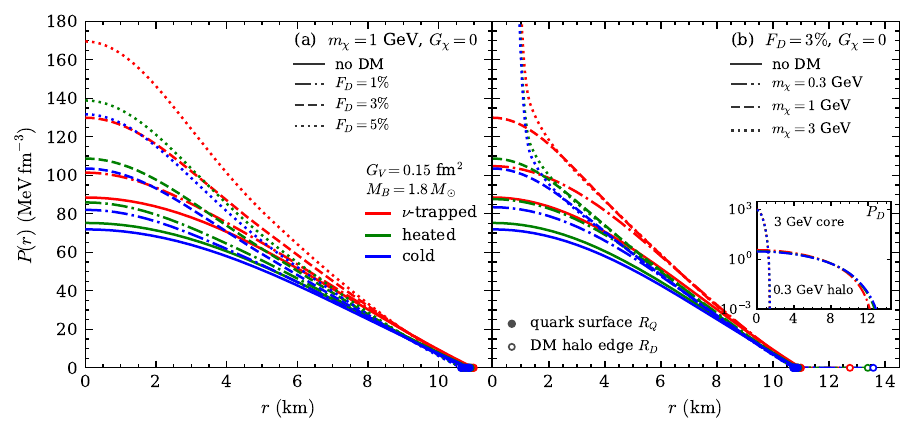}
\caption{Total pressure $P(r) = P_Q(r) + P_D(r)$ inside stars with $M_B = 1.8\,M_\odot$, constructed as in Fig.~\ref{fig:m_r}. (a)~The stars of Fig.~\ref{fig:m_r}(a); (b)~without DM and with $F_D = 3\%$ for $m_\chi = 0.3$, $1$, and $3$~GeV. Colors, line styles, and symbols as in Fig.~\ref{fig:m_r}. For $m_\chi = 3$~GeV the central pressure, $1.1$--$1.2 \times 10^3$~MeV\,fm$^{-3}$, lies above the range shown. The inset of panel (b) shows the DM pressure $P_D(r)$ alone for $m_\chi = 0.3$ and $3$~GeV on a logarithmic scale.}
\label{fig:P_r}
\end{figure*}
% ------------------------------

The pressure profiles are shown in Fig.~\ref{fig:P_r}, where panel (b) compares the three DM masses at $F_D = 3\%$. Without DM the central pressure decreases by about $20\%$ from the neutrino-trapped to the cold stage, while the central baryon density remains nearly constant; the decrease reflects the loss of the thermal and neutrino pressure. The central quark pressure of the stars with DM decreases by a similar fraction. The DM compresses the quark matter at the center. For $m_\chi = 1$~GeV and $F_D = 3\%$ the central quark pressure increases by about a third, and the DM adds only a small central pressure of its own. The compact $3$~GeV core has a central pressure of about $10^3$~MeV\,fm$^{-3}$, an order of magnitude above that of the star without DM, and more than doubles the central quark pressure. The $0.3$~GeV halo, in contrast, has a central pressure of only a few MeV\,fm$^{-3}$, which decreases over a large distance and remains finite at the quark surface (inset of Fig.~\ref{fig:P_r}).

% -------------------------------
% FIGURE - 10 
\begin{figure*}[tbp]
\centering
\includegraphics[width=\textwidth]{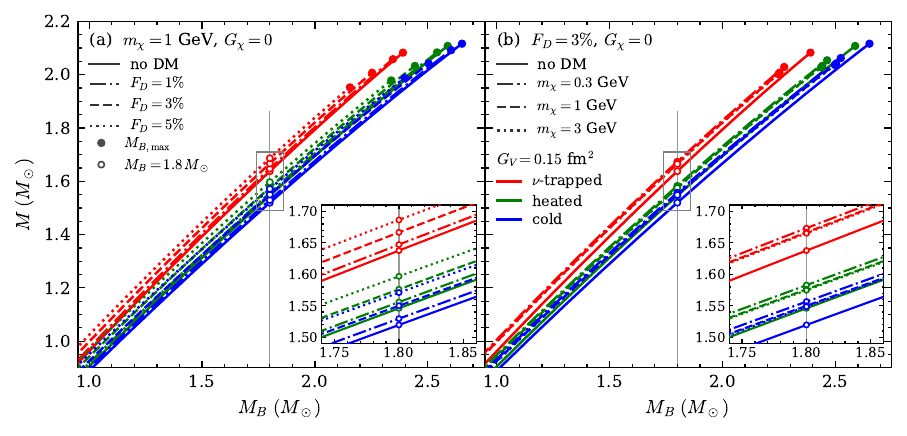}
\caption{Gravitational mass versus baryon mass in the neutrino-trapped (red), heated (green), and cold (blue) stages, for $G_V = 0.15$~fm$^2$ and $G_\chi = 0$. The DM curves are sequences at fixed DM rest mass, equal to that of the stars of Fig.~\ref{fig:P_r}. Panels and line styles as in Fig.~\ref{fig:P_r}. Filled dots mark the largest baryon mass $M_{B,\rm max}$ of each sequence. The vertical line is $M_B = 1.8\,M_\odot$, and the open circles on it are the stars of Fig.~\ref{fig:P_r}. The insets magnify the region around this line.}
\label{fig:M_MB}
\end{figure*}
% ----------------------------------
During the evolution a star moves vertically in the plane of Fig.~\ref{fig:M_MB}, at fixed baryon mass, from the neutrino-trapped to the cold curve. The figure shows both the energy released by a given star and whether the star remains stable. The largest baryon mass $M_{B,\rm max}$ of each sequence, given by the end points of the curves and listed in Table~\ref{tab:DMstages}, increases from the neutrino-trapped to the heated and cold stages in every case. A star that is stable at birth therefore remains stable throughout its early evolution, and no delayed collapse to a black hole occurs, with or without DM. The largest baryon mass that a newborn strange star can have is thus set by the neutrino-trapped stage. The DM lowers this limit, by $0.04$--$0.23\,M_\odot$ for $m_\chi = 1$~GeV and $F_D = 1$--$5\%$ (Table~\ref{tab:DMstages}), and the reduction grows with both the DM fraction and the DM mass, but the DM does not change the ordering of the stages. The same conclusion holds for $G_V = 0$ and for all other DM parameters, as we find for the $30$ combinations of $G_V$, $m_\chi$, $G_\chi$, and $F_D$ considered, with the DM rest mass of the cold maximum-mass star. This includes the three cases with $m_\chi = 3$~GeV and $F_D = 5\%$ in which the DM core is close to its own mass limit (Sec.~\ref{subsec:res_cold}). In these cases a star with the given $N_\chi$ exists only up to a certain central quark pressure, above which the DM core cannot be supported, and the maximum of $M_B$ lies slightly below this pressure. It is thus an ordinary turning point, to which the criterion of Eq.~(\ref{eq:stability}) applies as in the other cases. In the other $27$ cases, $M_{B,\rm max}$ increases by $0.12$--$0.20\,M_\odot$ during deleptonization and by $0.04$--$0.06\,M_\odot$ during cooling, and in the three cases close to the DM limit by $0.09$--$0.18\,M_\odot$ and $0.03$--$0.06\,M_\odot$, respectively.
% ----------------------------------
\subsection{Tidal deformability}
\label{subsec:res_tidal}
% -----------------------
% FIGURE - 11
\begin{figure*}[t]
\centering
\includegraphics[width=\textwidth]{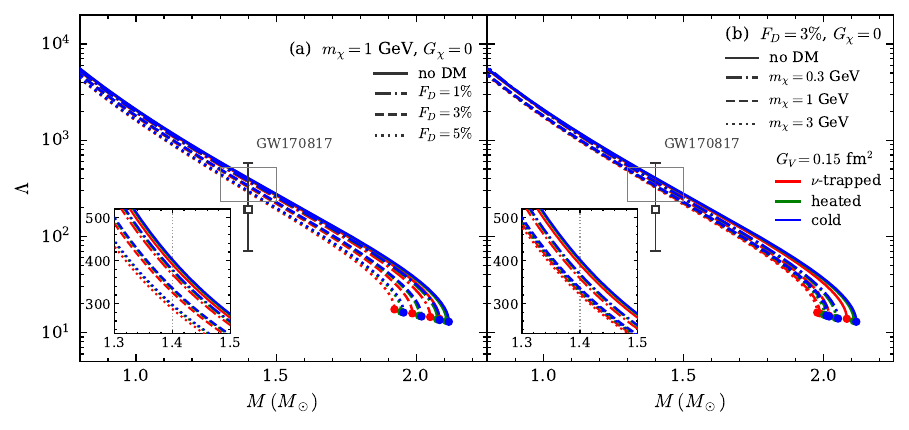}
\caption{Dimensionless tidal deformability as a function of the gravitational mass in the neutrino-trapped (red), heated (green), and cold (blue) stages, for $G_V = 0.15$~fm$^2$ and $G_\chi = 0$. (a)~$m_\chi = 1$~GeV with $F_D = 0$ (solid), $1\%$ (dash-dotted), $3\%$ (dashed), and $5\%$ (dotted); (b)~without DM (solid) and with $F_D = 3\%$ for $m_\chi = 0.3$ (dash-dotted), $1$ (dashed), and $3$~GeV (dotted), as in Fig.~\ref{fig:MR_stages}. Only the stable branches are shown, ending at the maximum masses (dots). The insets show the region around $M = 1.4\,M_\odot$ on a linear scale. The error bars show the constraint $\Lambda_{1.4} = 190^{+390}_{-120}$ ($90\%$ credibility) from GW170817~\cite{LIGOScientific:2018cki}, which applies to cold stars.}
\label{fig:Lambda}
\end{figure*}
% --------------------------------------------- 
The tidal deformability of the stars of Fig.~\ref{fig:MR_stages} is shown in Fig.~\ref{fig:Lambda}. We first discuss the cold stage, for which the effect of DM on $\Lambda$ has been studied before~\cite{PhysRevD.104.083016,Lopes:2023uxi,Yang:2024vjx}. Without DM we find $\Lambda_{1.4} = 404$ for $G_V = 0.15$~fm$^2$ and $271$ for $G_V = 0$. The DM reduces $\Lambda$, and the reduction grows with both the DM fraction and the DM particle mass, as for the radius (Table~\ref{tab:DMstages}). The reduction by $26\%$ for $m_\chi = 1$~GeV and $F_D = 5\%$ has two comparable sources: the smaller radius lowers the factor $C^{-5}$ by $12\%$, and the Love number decreases by $16\%$, because the DM core concentrates the mass toward the center. For $m_\chi = 0.3$~GeV the DM forms a halo, and $\Lambda$ is determined at its outer edge, $R = R_D$ (Sec.~\ref{subsec:tidal}). The larger radius would increase $\Lambda$ strongly, but the halo contains little mass, and the correspondingly small Love number nearly compensates for it, so that $\Lambda_{1.4}$ is reduced only slightly. The self-interaction changes $\Lambda_{1.4}$ by at most about $1\%$. All these values are compatible with the constraint from GW170817~\cite{LIGOScientific:2018cki}, so that the tidal deformability does not exclude any of the DM parameters considered here. This comparison is approximate, since GW170817 constrains the combined tidal deformability of the binary rather than $\Lambda_{1.4}$ itself.
% ----------------------------

In the hot stages the tidal deformability is slightly smaller than in the cold stage, by about $3\%$ in the neutrino-trapped stage and $1\%$ in the heated stage (Table~\ref{tab:DMstages}). This follows from the slightly smaller radius, and hence larger compactness, of the hot stars at fixed mass for $G_V = 0.15$~fm$^2$ (Fig.~\ref{fig:MR_stages}). The stage dependence is about ten times weaker than the effect of the DM, and the reduction of $\Lambda$ by the DM is nearly the same in all stages, $26$--$27\%$ for $m_\chi = 1$~GeV and $F_D = 5\%$, in line with the results of Sec.~\ref{subsec:res_stages}. Proto-strange quark stars are not observed in binary mergers, so these values have no direct observational counterpart. They show, however, that at fixed mass the tidal deformability of a DM-admixed strange star changes by only a few percent between the stages, whereas the DM changes it much more strongly.
% -----------------------------
\section{Summary and conclusions}
\label{sec:conclusions}
We have studied strange quark stars admixed with fermionic dark matter in the three stages of their early evolution: the neutrino-trapped stage, the heated stage after deleptonization, and the cold stage. Quark matter was described by the MIT bag model at finite temperature with a repulsive vector interaction, and the DM by a cold, asymmetric gas of fermions, with and without a repulsive self-interaction, that couples to the quark matter only through gravity. The two components were treated in the two-fluid formalism, from which we obtained the structure, the maximum masses, and the tidal deformability of the stars in each stage. To follow individual stars through the evolution, we compared configurations with the same numbers of baryons and DM particles. To our knowledge, DM-admixed strange quark stars have not been studied beyond the cold stage before, and such a star has not been followed at separately conserved numbers of baryons and DM particles through its early evolution.
% ---------------------------

Our main result concerns the stability of a newborn star. The largest baryon mass increases from the neutrino-trapped to the heated and cold stages in all cases considered, with and without DM, and also when the DM core is close to its own mass limit. A DM-admixed strange star that is stable at birth therefore remains stable during its early evolution, and the DM does not cause a delayed collapse to a black hole. The DM does, however, lower the largest baryon mass that a newborn star can have by $0.04$--$0.23\,M_\odot$ for DM fractions of $1$--$5\%$ with $m_\chi = 1$~GeV, so that the range of baryon masses for which a stable strange star can be formed becomes narrower.
% ----------------------

The effect of the DM on the global properties of the star is nearly the same in the three stages. For $m_\chi = 1$~GeV and $F_D = 5\%$, the DM lowers the maximum mass by about $8\%$ and the tidal deformability of a $1.4\,M_\odot$ star by about $26\%$ in each stage, whereas the stage itself changes these quantities by only a few percent. This is a consequence of the cold DM, whose EOS is the same in all stages, so that the stage enters only through the quark EOS. The size of the effect is set by the DM particle mass and fraction. Heavy DM forms a compact core that compresses the quark matter and reduces the maximum mass, the radius, and the tidal deformability most strongly, whereas light DM with $m_\chi = 0.3$~GeV can form an extended halo, which turns into a core in heavier stars when the DM fraction is small. In none of the cases does the DM increase the maximum mass. The self-interaction of the DM matters only for heavy DM. Without it, DM with $m_\chi = 3$~GeV reaches its own mass limit at $F_D \gtrsim 3$--$4\%$, and the maximum mass of the star is then set by the DM component rather than by the quark matter.
% --------------------------

During the evolution of a given star the DM keeps its particle number and does not take part in the cooling. The energy carried away by the neutrinos, about $0.12\,M_\odot$ for a star with $M_B = 1.8\,M_\odot$, is changed by the DM by less than $3\%$. Because the quark matter loses energy while the DM does not, the DM fraction of the star increases slightly as it evolves, and the DM distribution expands as the gravitational field of the quark matter weakens. Within our model, the observed mass of PSR~J0740$+$6620 requires a stiff quark EOS and limits the DM fraction to a few percent unless the DM is light, whereas the tidal deformability inferred from GW170817 does not exclude any of the DM parameters considered.
% ------------------------------------------

Our results are subject to several approximations. The early evolution was represented by a sequence of quasistatic configurations with prescribed entropy per baryon and lepton fraction, rather than by a solution of the neutrino transport equations. The DM was assumed to be cold and to interact with the quark matter only through gravity, so that neither heating of the DM nor energy exchange between the two components was included, and the DM fraction, whose astrophysical value is uncertain by many orders of magnitude, was treated as a free parameter. The quark matter was described by the bag model without perturbative QCD corrections, and the stars were taken to be bare, without a nuclear crust. Finally, we determined the stability limits from the maximum of the baryon mass at fixed particle numbers, i.e., from the vanishing of the Jacobian in Eq.~(\ref{eq:stability}), which marks the onset of instability of the fundamental radial mode~\cite{Caballero:2024qtv,Chen:2025soc,Kain:2020zjs}; we did not, however, solve the coupled radial-oscillation eigenvalue problem itself. A treatment of the neutrino transport in the presence of DM, a calculation of the radial modes of the two-fluid configurations, in particular near the limit of the DM core, and the inclusion of rotation are natural extensions of this work.
% ------------------------------------- 
\begin{acknowledgments}
R.S. gratefully acknowledges the Institute of Physics, Bhubaneswar, India, for support and research facilities during her visit.
\end{acknowledgments}
% ------------------------------------------
\bibliography{bibtex}
% ---------------------------------
\end{document}